\documentclass[twocolumn,showpacs,showkeys]{revtex4}

\usepackage{hyperref}
\usepackage{amsmath}
\usepackage{amsfonts}
\usepackage{amssymb}
\usepackage{graphicx}
\usepackage{graphics}

\usepackage{pstricks,pst-node,pst-plot} 

\begin{document}
\title{Analytical calculation of fragmentation transitions in adaptive networks}
\author{Gesa A. \surname{B\"ohme}}
\email{gesa@pks.mpg.de}
\affiliation{Max-Planck Institute for the Physics of Complex Systems, N\"othnitzer Str. 38, 01187 Dresden, Germany}
\author{Thilo \surname{Gross}}
\affiliation{Max-Planck Institute for the Physics of Complex Systems, N\"othnitzer Str. 38, 01187 Dresden, Germany}
\date{\today}
\begin{abstract}
In adaptive networks fragmentation transitions have been observed in which the network breaks into disconnected components.
We present an analytical approach for calculating the transition point in general adaptive network models. Using the example of an adaptive voter model, we demonstrate that the proposed approach yields good agreement with numerical results. 
\end{abstract}
\maketitle{}
In the past decade networks have proven to offer a metaphor for describing and analyzing complex systems in different fields 
with applications ranging from technological to biological and social systems \cite{Barabasi,dorogovnew,NewmanRev,Bocca}.
By conceptually reducing a complex system to a set of discrete \emph{nodes} connected by \emph{links} a simplification is achieved that often 
enables deep insights in the structure and dynamics of the system. 

In network physics, dynamics can refer to two different concepts.
First, the dynamics \emph{OF} networks describes the temporal evolution of the topology, the specific pattern of nodes and links. Second, the dynamics \emph{ON} networks refers to the evolution of internal properties in the network nodes, which are coupled according to the (typically static) topology.
Systems combining both types of dynamics are called adaptive or coevolutionary networks \cite{anbook,ThiloAdap}.  

Adaptive networks are presently studied in several different disciplines because they appear in many different applications \cite{wiki} and exhibit some unique phenomena, including robust self-organization to critical states \cite{BornholdtRohlf}, emergence of distinct classes of nodes from an initially homogeneous population \cite{Kaneko}, formation of complex hierarchical topologies \cite{SanMiguel, zimmermann2000} and complex network-level dynamics and phase transitions \cite{GrossDommarBlasius}. 

One exciting recent discovery in the field of adaptive networks is the existence of a generic scenario for fragmentation transitions \cite{VazquezGeneralVM}.
In many adaptive networks, dynamics is contingent on the presence of so-called \emph{active links} that connect nodes in different states.
When no such links are left, dynamics stops and an absorbing state is reached.
Absorbing states are therefore encountered either when the network becomes polarized so that all nodes are in the same state or when the network fragments such that nodes separate into disconnected components, which are internally state-uniform.
In the following we denote the phase transition that separates the fragmented phase from the polarized phase as the fragmentation transition. 

Fragmentation transitions are frequently observed in simulations \cite{HolmeNewman,KozmaBarrat,GilZanetteAdaptVM,VazquezGeneralVM,Kimura}.
However, for a detailed understanding of dynamics in more complex future models analytical approaches to the fragmentation transition will be instrumental.
Existing approaches that faithfully capture other transitions \cite{GrossDommarBlasius,shaw,gerd,gusman,marceau,VazquezPcNodeup, Kimura, Guven} yield only very rough approximations for the fragmentation threshold. A paradigmatic example is provided by the adaptive voter model that describes opinion formation in networked populations \cite{HolmeNewman,KozmaBarrat,GilZanetteAdaptVM,VazquezPcNodeup,Kimura,Guven}.
In this model mean-field and pair approximations overestimate the fragmentation threshold by 150-200\% \cite{VazquezPcNodeup, Kimura, Guven}.

In the present paper we propose a simple analytical approach for determining fragmentation thresholds in adaptive networks.
This approach takes inspiration from percolation arguments that are commonly used in the computation of the epidemic threshold of infectious diseases.
Below we start by providing a brief overview of the adaptive voter model, which is subsequently used as an illustrative example. We then propose a first approach for computing the fragmentation threshold in networks. This approach relies on the assumption that 
the degree distribution, i.e. the distribution of the number of links connecting to the network nodes, is sufficiently narrow. 
Finally, we extend this approach to arbitrary degree distributions.     

The voter model was originally proposed as a conceptual model of opinion formation \cite{DefinitionVM}. The model considers a network in which 
the nodes represent social agents and the links represent social contacts. Each agent can hold either of two symmetric opinions. 
The opinions are updated continuously by 
either a) selecting a random agent and letting it adopt the opinion of a randomly chosen neighbor (direct voter model), b) selecting a random agent and copying its 
opinion to a randomly chosen neighbor (reverse voter model) or c) selecting a random link and letting one of the linked agents adopt the others opinion (link-update voter model). In the non-adaptive variants of the voter model the network topology remains static. If the network is connected the dynamics 
therefore continues until eventually an absorbing consensus state is reached in which all nodes hold the same opinion.

Adaptive extensions of the voter model take an additional process into account. At a certain rate agents that are connected to 
agents of different opinion break the respective link and connect to a randomly chosen agent of their own opinion. This rewiring can lead to a fragmentation of the network, such that both opinions survive in disconnected network components that are internally in consensus \cite{Nardini,vazquez2007,GilZanetteAdaptVM,HolmeNewman}.
Below, we consider an adaptive voter model with link update in which rewiring events occur at a rate $p$ and opinion updates occur at a rate $1-p$. 
The challenge in this model is to compute the fragmentation threshold, i.e. the critical value of $p$ at which the active links disappear.  

To simplify the discussion below we define the notion of a $q$-fan motif: a subgraph consisting of one node holding a given opinion and $q$ neighbors of the node 
holding the other opinion. The node holding the solitary opinion is denoted as the \emph{base node} of the fan, whereas the other nodes are denoted as the \emph{fringe nodes}. For instance a 4-fan contains a base node and 4 fringe nodes, which are connected to the base node by active links.

For computing the fragmentation transition we consider a single active link connecting two almost disconnected clusters of different opinions. 
We then study the dynamics in the vincity of this link and compute the net balance of active links. If this balance is negative then the number of active links 
declines exponentially leading eventually to the fragmented absorbing state. If the balance is positive then active links proliferate preventing the network 
from reaching the fragmented state. 

\begin{figure}
 \begin{center}
 \includegraphics[trim=6cm 18cm 0 5cm,width=9cm]{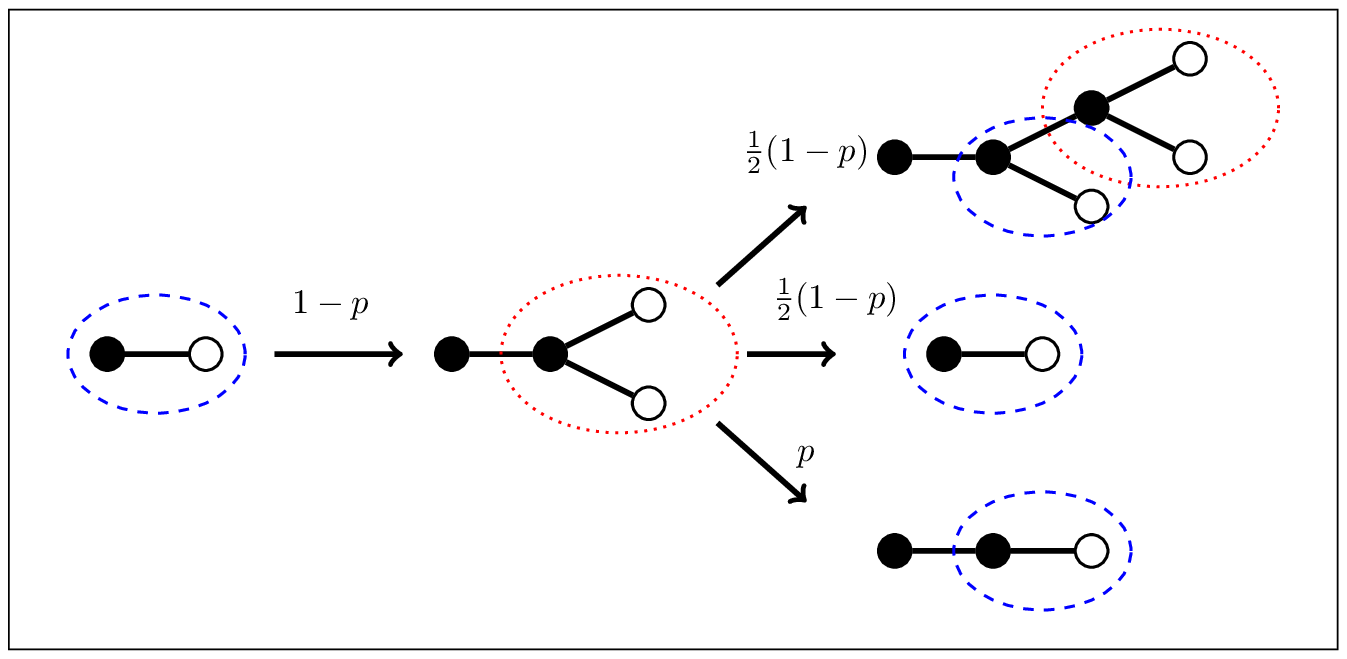}
\caption{(Color online) Illustration of the evolution of active links in a degree regular network with degree $k=3$.
Agents are depicted as nodes which are open or solid depending on their opinion. Shown is the network 
in the neighborhood of an active link connecting nodes of different opinions. Arrows correpond to dynamical 
updates and are labeled with the corresponding transition rate. Depending on the parameters the updates lead to 
proliferation or decline of isolated active links (encircled dotted) and 2-fan motifs (encircled 
dashed).\label{bundleappr}}
 \end{center}
\end{figure}

For simplicity let us first assume that the network is \emph{degree regular}, so that every node has excactly the same number of neighbors $k$.
For the special case of $k=3$ the dynamics of active links is illustrated in Fig.~\ref{bundleappr}. We start from a single active link (left half of figure). 
In the next update that affects this link, the link will be rewired with probability $p$ becoming inert (not shown), with the complementary 
probability $1-p$ one of the nodes connected by this link adopts the other's opinion. In the adoption event the original active link becomes inert, but
the $k-1$ other connections of the adopting agent become active.
   
The active links that are created by an adoption event form a $k-1$-fan.   
Because these links can become inert simultaneously if the node at the base of the fan changes its opinion, the links of the fan cannot be considered 
independent. We therefore continue by studying how the next update alters the $k-1$-fan motif (Fig.~\ref{bundleappr} right half). 
If the update is a rewiring event (probability $p$) then it decreases the width of the fan turning the $k-1$-fan into the $k-2$-fan. If the update is an opinion adoption event (probability $1-p$) 
then there are two possible scenarios occuring with equal probability. In the first scenario the node at the base of the 
fan changes its opinion. In this case the whole $k-1$-fan becomes inert, but one new active link is formed at the base of the   
fan. In the second scenario one of the fringe nodes of the fan adopts the base node's opinion, in this case the width of the 
fan is reduced by one, but an additional $k-1$-fan is activated.   

\begin{figure}
 \begin{center}
 \includegraphics[trim=.5cm 2cm 0 0,width=9cm]{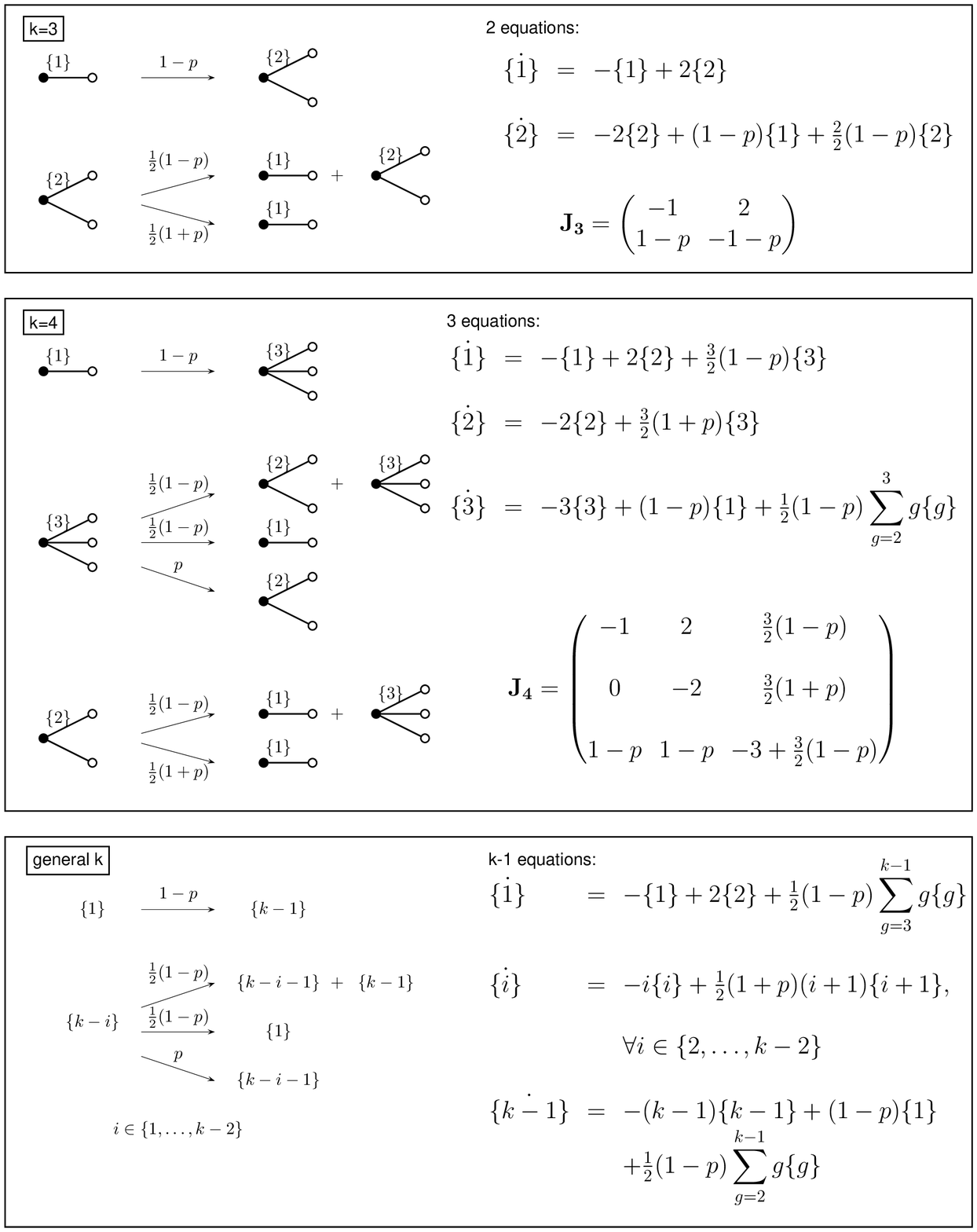}
\caption{Illustration of the transitions for the $q$-fan motif $\{q\}$, in degree regular networks with degree $k$.
Motifs and transitions are depicted as in Fig. 1. Additionally, the corresponding systems of differential equations and the corresponding 
Jacobians are shown (see text).}
\label{motifs}
 \end{center}
\end{figure}
 
For $k=3$, a $k-2$-fan is just a single active link. We thus obtain a closed system of transitions. Summarizing the results of Fig.~\ref{bundleappr}, we find that an update affecting a single active link, always deactivates the link and activates a 2-fan with probability $1-p$. An update affecting a 2-fan always turns the  2-fan into a single active link and additionally creates one 2-fan with probability $(1-p)/2$ (Fig.~\ref{motifs} top panel). These observations can be summarized further in a system of differential equations decribing the motifs
\begin{equation}
\begin{array}{r c l}
 \dot{\{1\}}&=&-\{1\}+2\{2\}\\
 \dot{\{2\}}&=&-2\{2\}+(1-p)\{1\}+2\cdot\frac{1}{2}(1-p)\{2\},
\end{array}
\end{equation}
where $\{q\}$ denotes the density of $q$-fans. Because we assumed that the density of active motifs is low the equations of motion are linear and 
can therefore be solved straightforwardly. In particular the fragmented state $\{ 1\} =\{ 2\} =0$ is stable and globally attractive if all eigenvalues 
of the corresponding Jacobian
\begin{equation}
 {\rm \bf J}=\begin{pmatrix}-1&2\\1-p&-1-p\end{pmatrix},
\end{equation}
have negative real part. The phase transition of the network thus corresponds to a bifurcation of the low dimensional model in which an eigenvalue
acquires a positive real part. For $k=3$ this transition occurs at $p=1/3$.   

Degree regular networks with $k>3$ can be treated analogously to the $k=3$ example. The corresponding equations for the $k=4$ and the generalization to arbitrary $k$ is shown in Fig.~\ref{motifs}. An update affecting an active link deactivates the link and activates a $k-1$-fan with probability $1-p$. An update affecting a $q$-fan either a) deactivates the fan and activates a single link (probability $(1-p)/2$); b) decreases the width of the fan by 1, turning the $q$-fan into a $q-1$-fan (probability $p$); or c) decreases the width of the fan by one and activates a new $k-1$-fan (probability $(1-p)/2$). This leads to a system of $k-1$ linear differential equations and the corresponding Jacobian. From this Jacobian the transition points can then be computed either by computation of eigenvalues or by direct analytical construction of testfunctions as described in \cite{GrossPhysicaD}. 

\begin{figure}
\begin{center}
  \includegraphics[width=7cm,trim=0 1cm 0 0,angle=270]{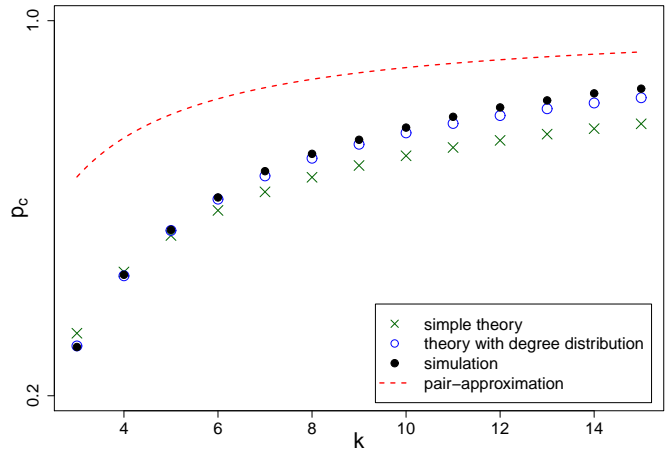}
\caption{(Color online) Fragmentation threshold in the adaptive voter model. Shown are numerical results from agent based simulation (black dots), pair-approximation (dashed line), the analytical approach proposed here (crosses), and its refined variant (circles). The proposed approaches yield a better match with the numerical results than the established procedure. Numerical simulations used $N=10^6$ nodes. In the refined approximation a cut-off of $k_{\max}=30$
was used.}
\label{compare}
\end{center}
\end{figure}

The assumption of degree regularity is questionable because even if present in the initial network, degree regularity is destroyed by adaptive rewiring.
Nevertheless, the comparison in Fig.~\ref{compare} shows that the procedure proposed above yields significantly better results than a common pair approximation.
For obtaining a closer match with simulation results we propose a refined procedure that takes the effect of broader degree distributions into account. 
Instead of the fan motif used above, we now consider \emph{spider} motifs, containing one central base node which is connected to $m$ nodes of it's own opinion and $l$ nodes of opposing opinion. The $\{m,l\}$-spider thus has $m$ inert links and $l$ active links, leading to a total degree $d=m+l$. As before we only consider active spiders, i.e. $l>0$.

\begin{figure}
 \begin{center}
 \includegraphics[trim=4cm 21cm 0 1cm,width=10cm]{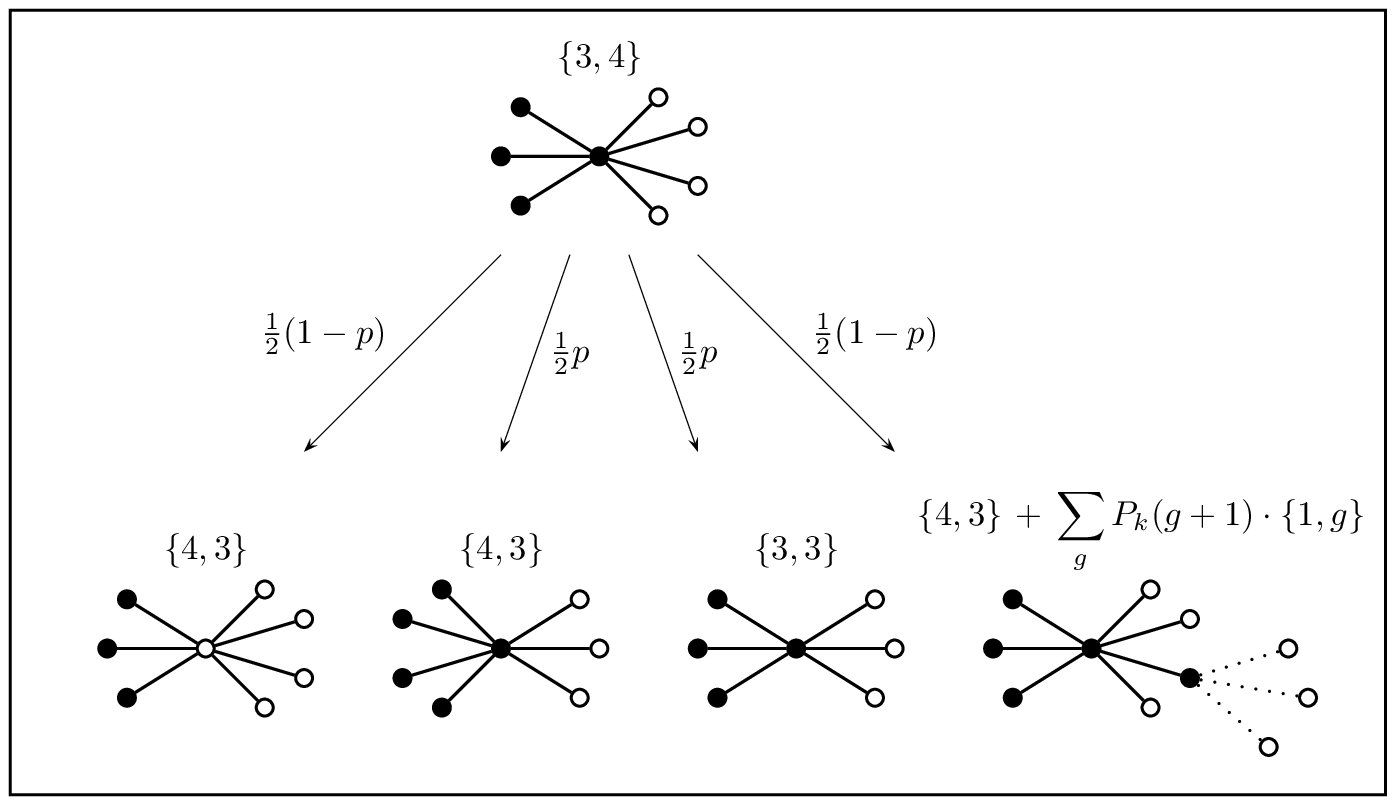}
\caption{Example for the transition probabilities of a $\{3,4\}$-spider motif. Diagrams and arrows represent motifs and transitions in analogy to Fig.~2.
The dashed links indicate a number of second neighbors which is drawn from a distribution (here, Poisson distribution).}
\label{deg34}
 \end{center}
\end{figure}

The effects of updates on a spider motif are shown schematically in Fig.~\ref{deg34}. In a rewiring event either the rewired link is kept by the fringe node and the $\{m,l\}$-spider is turned into a $\{m,l-1\}$-spider or the rewired link is kept by the base node turning the $\{m,l\}$-spider into a $\{m+1,l-1\}$-spider. In an opinion adoption event either the base node is convinced, which turns all active links into inert links and vice versa, leading to a $\{l,m\}$-spider, or one of the fringe nodes is convinced by the base node giving rise to a new $\{1,g\}$-spider while in the remaining spider one active link turns into an inert link. In the latter case $g$, the number of active links of the newly activated spider is given by the next-neighbor degree distribution $P(g+1)$. In the adaptive network this distribution is reshaped by the rewiring events. The exact distribution that is observed in the long-term behavior is therefore not known. However, the relatively good results obtained already with the degree-regular approximation suggest that a precise knowledge of the distribution is not necessary. In the present example we therefore use a Poisson distribution with mean degree $k$, $P(x)=e^{-k}k^{x}/x!$, which adequately captures the effect of random rewiring.
The transition rules can then be summarized as
\begin{align*}
\{m,l\}\left\{
\begin{array}{cl}
 \xrightarrow{(1-p)/2}&\{l,m\}\\
\xrightarrow{(1-p)/2}&\{m+1,l-1\}+\displaystyle\sum_{g=1}^{k_{max}-1} \frac{e^{-k}k^{g+1}}{(g+1)!}\cdot \{1,g\}\quad\qquad\\
\xrightarrow{p/2}\,&\{m+1,l-1\}\\
\xrightarrow{p/2}\,&\{m,l-1\}.
\end{array}
\right.
\end{align*}
For obtaining a Jacobian of finite size we consider only motifs with $k=m+l \leq k_{\rm max}$. 
Following the same steps as above this leads to a $(D\times D)$ Jacobi matrix, where $D=k_{\rm max}(k_{\rm max}+1)/2$. 
Because of its complexity we construct this matrix by computer algebra and compute the fragmentation threshold by numerical eigenvalue computation.  

The comparison in Fig.~\ref{compare} shows that the results from the refined procedure are in excellent agreement with simulation results. 
In particular the approximation works well for low $ k $ where pair and mean-field expansions yield poor results. For high $ k $
there is a small discrepancy, which may be due to long-range correlations or the precise shape of the degree distribution.  
 
In summary we proposed two approaches for computing fragmentation thresholds in adaptive networks. The simpler approach allows for a quick analytical estimation 
of the threshold with higher accuracy than previously proposed approaches. The refined approach yields numerical predictions with high accuracy. 
In contrast to other techniques the proposed approaches lead to linear ODEs because they assume a low density of active links. While the linear ODEs 
can be solved exactly for any density of active links, the approximation must fail when the system is far from the fragmentation point.
A difficulty not addressed in the present paper is obtaining the degree distribution for an adaptive network without explicit simulation.
In principle the relevant information for applying the refined procedure proposed here could be obtained by third-order moment expansions \cite{Kimura}, which 
cannot predict the fragmentation transition precisely, but allow relatively faithful estimation of the width of the degree distribution. 
In practice it may be simpler to use statistical fits of distributions observed in a small set of exploratory simulation runs. Further, using a Poisson distribution should yield good results for all networks with exponentially decaying degree distributions. We therefore believe that both the simple and the refined procedure proposed here can be applied with relative ease in practice and will be instrumental to exploring fragmentation transitions in future adaptive network models. 

The authors thank G.~Demirel and F.~Vazquez for insightful discussions.


\end{document}